\documentclass[]{aa}  
\usepackage{graphicx}
\usepackage{txfonts}
\usepackage[]{hyperref}

\begin{document}

   \title{Search for helium in the upper atmosphere of the hot Jupiter WASP-127~b using Gemini/Phoenix}

   \author{Leonardo A. dos Santos
          \inst{1}
          \and
          David Ehrenreich
          \inst{1}
          \and
          Vincent Bourrier
          \inst{1}
          \and
          Romain Allart
          \inst{1}
          \and
          George King
          \inst{2, 3}
          \and
          Monika Lendl
          \inst{1}
          \and
          Christophe Lovis
          \inst{1}
          \and
          Steve Margheim
          \inst{4}
          \and
          Jorge Mel\'endez
          \inst{5}
          \and
          Julia V. Seidel
          \inst{1}
          \and
          S\'ergio G. Sousa
          \inst{6}
          }

\institute{Observatoire astronomique de l’Université de Genève, 51 
chemin des Maillettes, 1290 Versoix, Switzerland\\
\email{Leonardo.dosSantos@unige.ch}
\and Department of Physics, University of Warwick, Gibbet Hill Road, Coventry, CV4 7AL, UK
\and Centre for Exoplanets and Habitability, University of Warwick, Gibbet Hill Road, Coventry, CV4 7AL, UK
\and Gemini Observatory/NSF’s NOIRLab, Casilla 603, La Serena, Chile
\and
Universidade de S\~ao Paulo, Departamento de Astronomia do IAG/USP, Rua do Mat\~ao 1226,
Cidade Universit\'aria, 05508-900 S\~ao Paulo, SP, Brazil
\and Instituto de Astrof\'isica e Ci\^encias do Espa\c{c}o, Universidade do Porto, CAUP, Rua das Estrelas, 4150-762 Porto, Portugal
}

   \date{Received 1 July 2020; accepted 12 July 2020}
 
  \abstract
   {Large-scale exoplanet search surveys have shown evidence that atmospheric escape is a ubiquitous process that shapes the evolution and demographics of planets. However, we lack a detailed understanding of this process because very few exoplanets that have been discovered to date could be probed for signatures of atmospheric escape. Recently, the metastable helium triplet at 1.083 $\mu$m has been shown to be a viable window for the presence of He-rich escaping envelopes around short-period exoplanets. Our objective is to use, for the first time, the Phoenix spectrograph to search for helium in the upper atmosphere of the inflated hot Jupiter WASP-127~b. We observed one transit and reduced the data manually since no pipeline is available. We did not find a significant in-transit absorption signal indicative of the presence of helium around WASP-127~b, and we set a 90\% confidence upper limit for excess absorption at 0.87\% in a 0.75 \AA\ passband covering the He triplet. Given the large scale height of this planet, the lack of a detectable feature is likely due to unfavorable photoionization conditions for populating the metastable \ion{He}{I} triplet. This conclusion is supported by the inferred low coronal and chromospheric activity of the host star and the old age of the system, which result in a relatively mild high-energy environment around the planet.}


   \titlerunning{Search for helium in the upper atmosphere of WASP-127~b}
   \authorrunning{L. A. dos Santos et al.}

   \maketitle
%

\section{Introduction}

The results from surveys of transiting exoplanets are fundamental to obtain a global overview of the exoplanet population in the solar neighborhood. One of the most important discoveries of these surveys is that the population of giant planets displays a dearth of hot, Neptune-sized worlds \citep{2007A&A...461.1185L, 2016A&A...589A..75M}. One of the main explanations proposed for this feature is atmospheric loss: Planets start their lives with a thick envelope that is rich in volatiles, which is gradually eroded by the high-energy stellar irradiation \citep[e.g.,][]{2003ApJ...598L.121L}. However, very few exoplanets have been revealed to be losing their atmospheres \citep[e.g.,][]{2003Natur.422..143V, 2010A&A...514A..72L, 2012ApJ...760...79H, 2015Natur.522..459E, 2018A&A...620A.147B, 2019AJ....158...91S}, which sets a limit on how sophisticated escape models can be used to explain observational results.

Classical observations for studying atmospheric escape have been performed in the ultraviolet (UV), mostly using the transmission spectroscopy technique in the Lyman-$\alpha$ (\ion{H}{I}) line at 1215.67~\AA. The escape of heavier particles has also been detected in the UV around highly-irradiated, nearby hot-Jupiters \citep[e.g.,][]{2004ApJ...604L..69V, 2013A&A...553A..52B, 2019AJ....158...91S}. These UV observations can only be performed from space since the Earth's atmosphere is opaque to these wavelengths. Another limitation of Lyman-$\alpha$ transit spectroscopy is that the interstellar medium (ISM) efficiently absorbs the core of the line, so we can only access information at high Doppler velocities (i.e., above $\sim$30 km~s$^{-1}$, depending on the radial velocity of the host star). Furthermore, the Lyman-$\alpha$ emission line, including its wings, is completely absorbed for distances larger than $\sim$100 pc.

More recently, \citet{2018ApJ...855L..11O} have revised the prediction of \citet{2000ApJ...537..916S} that the metastable \ion{He}{I} triplet at 1.083 $\mu$m can be used in transmission to study atmospheric properties of transiting planets, including escape. This wavelength window is not affected by ISM absorption. To date, several studies have come out exploiting this new technique \citep[e.g.,][]{2018Natur.557...68S, 2018Sci...362.1384A, 2018ApJ...868L..34M, 2018Sci...362.1388N, 2018RNAAS...2...44K, 2019A&A...623A..58A, 2019A&A...629A.110A, 2020AJ....159..115K, 2020ApJ...894...97N, 2020MNRAS.495..650G, 2020arXiv200413728V}. Since there are more cases of planets that are amenable to infrared transit spectroscopy than in the UV, and as it is impossible to probe all of them with the few instruments currently available, the community would benefit from including Phoenix in the roster of those capable of conducting \ion{He}{I} transmission spectroscopy. 

Our objective in this study is to search for He in the upper atmosphere of the inflated hot Jupiter WASP-127~b \citep[][]{2017A&A...599A...3L} and assess the capability of an alternative spectrograph in the Southern Hemisphere for this purpose. The orbital and planetary parameters of WASP-127~b were recently updated with a joint analysis of \emph{TESS} and Euler photometry as well as CORALIE radial velocities, yielding a radius of $R_{\rm pl} = 1.311^{+0.025}_{-0.029}$ R$_{\rm Jup}$, a mass of $M_{\rm pl} = 0.165^{+0.021}_{-0.017}$ M$_{\rm Jup}$, and an orbital semi-major axis of $a_{\rm pl} = 0.0484^{+0.0013}_{-0.0009}$ au (Seidel et al., in prep.). The planet has recently been shown to display a feature-rich transmission spectrum from near-UV to near-infrared wavelengths \citep{2018A&A...616A.145C, 2019arXiv191108859S}, thus bringing further evidence that WASP-127~b is a promising target for atmospheric characterization.

\section{Observation and data reduction}

We observed one transit of WASP-127~b in the night of 17 March 2019 by using the visiting instrument Phoenix spectrograph \citep{2003SPIE.4834..353H} installed on the Gemini South Telescope (PI: dos Santos, Director's discretionary time program GS-2019A-DD-105). This observation was a pilot study to assess the feasibility of using the Phoenix spectrograph to measure He extended atmospheres around transiting exoplanets. We used the J9232 slit with an aperture of 0.25 arcsec (three pixels), resulting in a nominal resolving power of $R \sim\ $65,000 near the He triplet. The exposure times were set to 900~s and we obtained 24 spectra covering the transit and the out-of-transit baseline. Two in-transit exposures had to be discarded because of high noise levels, possibly caused by the passage of clouds.

The Phoenix spectrograph is neither cross-dispersed nor stabilized like other high-resolution spectrographs that are currently used for cutting-edge exoplanet science; this lack of wavelength stability does not, however, preclude us from using it to conduct transmission spectroscopy studies. Phoenix is capable of spectroscopy across the $1 - 5$ $\mu$m region, and the wavelength coverage is limited to the bandpass of the blocking filter mounted on the instrument and the finite size of the detector. The spectrograph does not possess an automatic data reduction pipeline, so we performed the reduction manually using \texttt{IRAF} and the packages \texttt{CCDPROC} and \texttt{CCDRED}. Besides the science exposures, the full data set also consists of 11 dark exposures, 20 flat-field exposures, and eight telluric standard exposures. The uncertainties of all exposures were calculated assuming a gain of 9.2 electrons per analog-to-digital unit (ADU) and a readout noise of 40 electrons per pixel. All of the uncertainties quoted in this study were calculated by propagating the uncertainties of the raw data in all of the steps of the data reduction and analysis process.

All exposures were trimmed in the wavelength dispersion direction, removing the first 100 pixels and the last 24 pixels to avoid noisy regions. The dark and flat-field frames were combined as a median with 3$\sigma$ clipping to produce the master dark and flat-field frames. We subtracted the dark from the master flat-field and normalized it by its overall mean to obtain the response function of the detector.

\begin{figure}[h]
\includegraphics[width=\hsize]{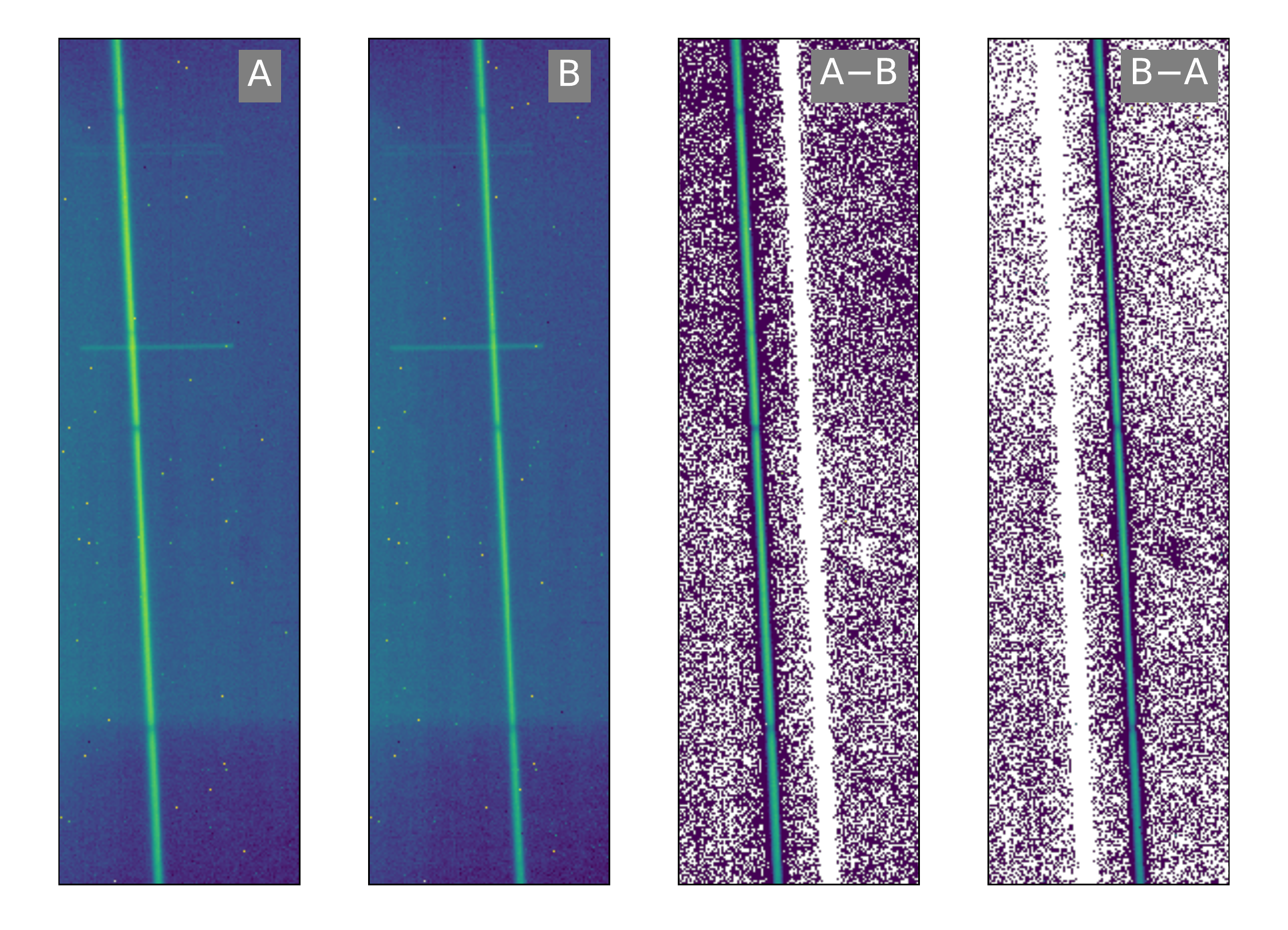}
\caption{Two consecutive science exposures (left) with spectra in different positions were subtracted of each other to remove the sky background (right).}
\label{sky_sub}
\end{figure}

The levels of sky background are highly variable near 1 $\mu$m. The strategy to perform sky subtraction with Phoenix observations consists in alternating the position of the stellar spectrum in the detector between each exposure. These positions are named A and B, and consecutive observations follow the pattern ABBA; each pair AB or BA is subtracted of each other's exposure, resulting in images without a sky and dark-current background (see Fig. \ref{sky_sub}). All telluric and science frames are then divided by the normalized master flat-field image.

The spectra are finally extracted using \texttt{IRAF}'s task \texttt{APALL} from each exposure. The only manual step is to set the initial aperture location in the spatial direction. This is done by finding the center of the stellar profile in two cross-dispersion cuts; one is in the upper and one is in the lower part of the detector. We then fit the spectral trace with a third-order polynomial and two iteration steps using the two previously mentioned points as a starting point for the fit. The width of the spectral trace is calculated automatically by the \texttt{APALL} task. We extracted the spectra for the science and telluric standard exposures using a regular box extraction algorithm.

We performed the wavelength calibration using the \texttt{IRAF}'s tasks \texttt{IDENTIFY} and \texttt{DISPCOR}; as a reference, we used the telluric lines available in the wavelength range of the spectra (1080.76-1084.62 nm). The average root mean square (RMS) of the wavelength calibration is $\sim$0.001 nm per pixel or $\sim$0.28 km~s$^{-1}$ in Doppler space, which represents the internal precision of the wavelength calibration. We measured the wavelength stability of Phoenix near the 1.083 $\mu$m He triplet and found that it is, on average, $\sim$1.6 pixels or $\sim$2.2~km~s$^{-1}$ in Doppler space. For comparison purposes, the Doppler trajectory of the planet WASP-127~b during transit varies between approximately $-30$ and $+30$~km~s$^{-1}$. Each spectra of WASP-127 is corrected for the barycentric radial velocity calculated with \texttt{barycorrpy}, a Python implementation of the algorithm by \citet{2014PASP..126..838W}, and then shifted to the stellar rest frame. The last step of the data reduction is to normalize the spectra of WASP-127; this is achieved by fitting the continuum of the spectra with the task \texttt{CONTINUUM}. For the sake of consistency, we fit the continuum of all spectra with three cubic spline pieces. The average signal-to-noise ratio (S/N) of the spectra near the He triplet is $\sim$26. In the beginning of the observation, there is no telluric contamination in the wavelength range of interest near the He triplet (see Fig. \ref{oot_spec}). But we found a feature during the transit near 1083.35~nm that could be caused by telluric lines (see Fig. \ref{time_series}). This feature is, however, far from the wavelength range of the \ion{He}{I} triplet, so there was no need to correct for it in our analysis.

\section{Results and discussion}

\begin{figure}[h]
\includegraphics[width=\hsize]{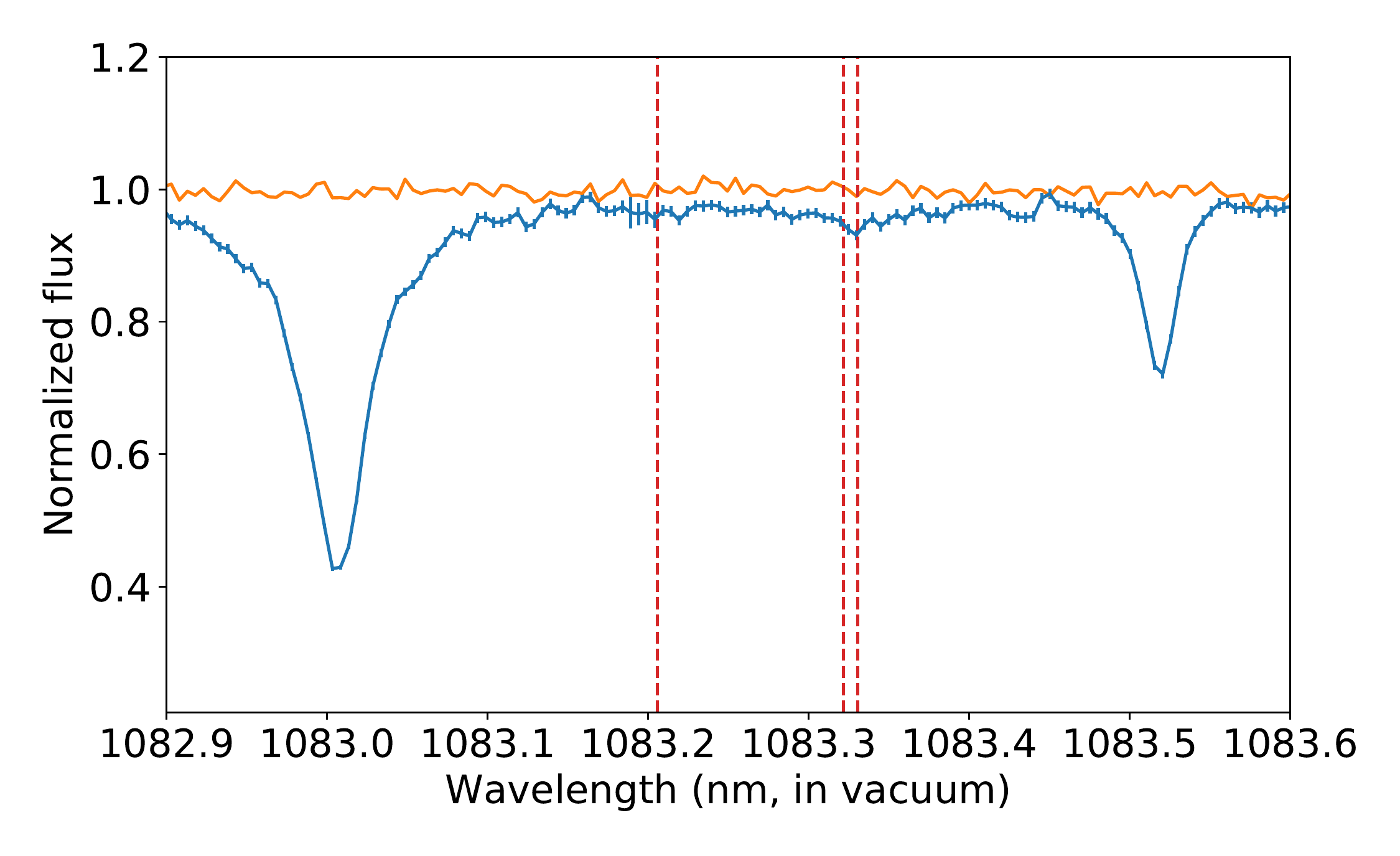}
\caption{Out-of-transit spectrum of WASP-127 around the He triplet (blue) and a telluric standard spectrum for reference (orange). The vertical-dashed lines represent the central wavelengths of each line in the triplet.}
\label{oot_spec}
\end{figure}

Since the spectra are normalized, we lose information about the flux decrease by the opaque disk of WASP-127~b; our analysis is thus only sensitive to excess absorption caused by the planet's atmosphere at specific spectral lines. In total, we obtained ten spectra outside the transit, and we combined them to produce a master out-of-transit spectrum (see Fig. \ref{oot_spec}). Contrary to previously observed targets, such as WASP-107 and HAT-P-11, the host star WASP-127 does not exhibit strong absorption lines of He in its spectrum. However, this does not preclude us from searching for an in-transit signal produced by the transiting planet since it would produce absorption in the stellar continuum instead. In fact, we expect that it improves the search for planetary He by increasing the local S/N (more photons than if there was a stellar line) and by avoiding flux variations due to the shifting stellar line in the planet rest frame.

\begin{figure}[h]
\includegraphics[width=\hsize]{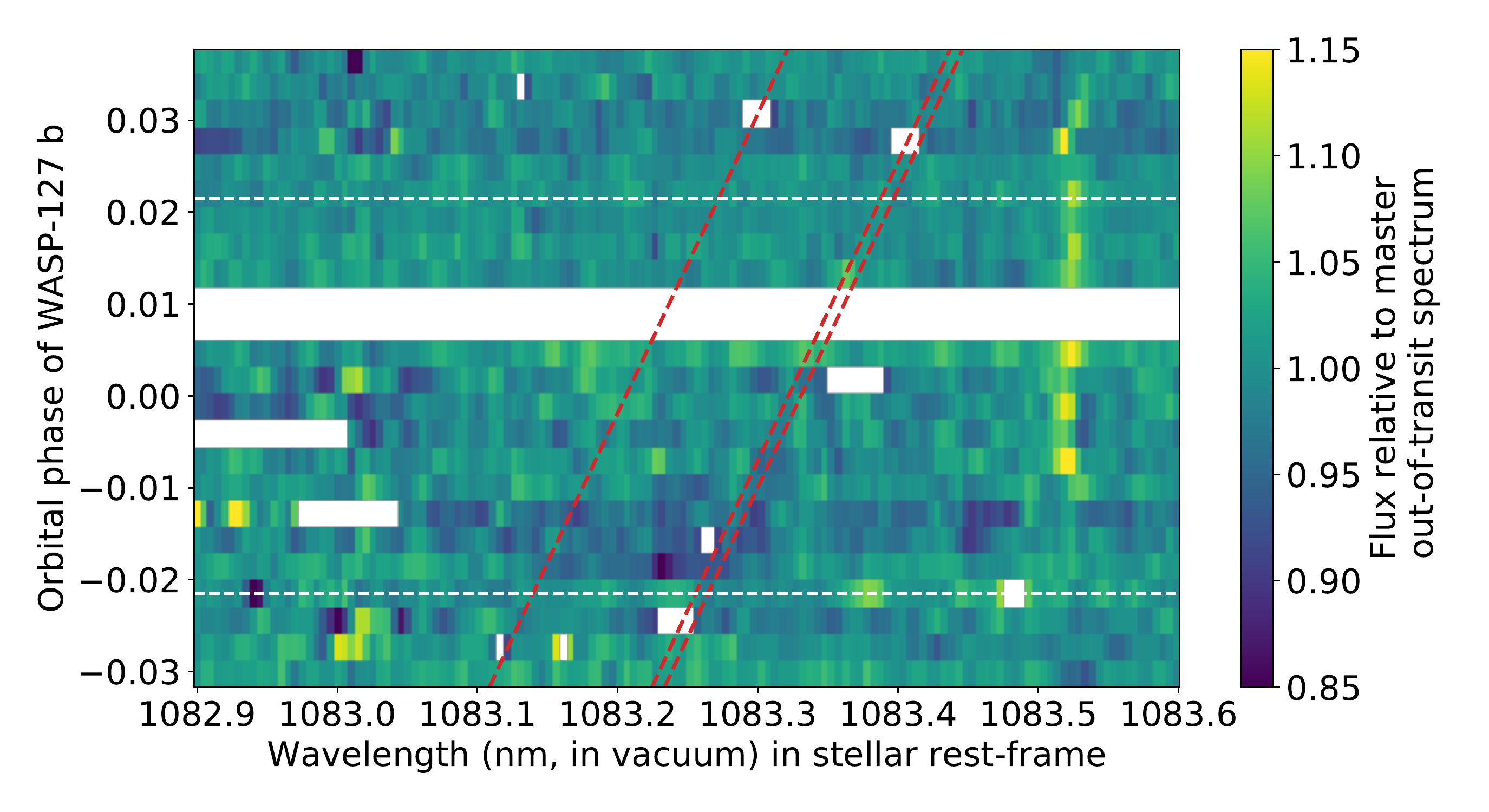}
\caption{Time series of the ratio between the individual spectra and the master out-of-transit spectrum; absorptions are positive and white regions represent bad pixels or discarded exposures. The horizontal-dashed lines represent the phases of ingress and egress of WASP-127~b. The dashed-red lines represent the central wavelength of the He triplet in the planetary rest frame.}
\label{time_series}
\end{figure}

\begin{figure}[h]
\includegraphics[width=\hsize]{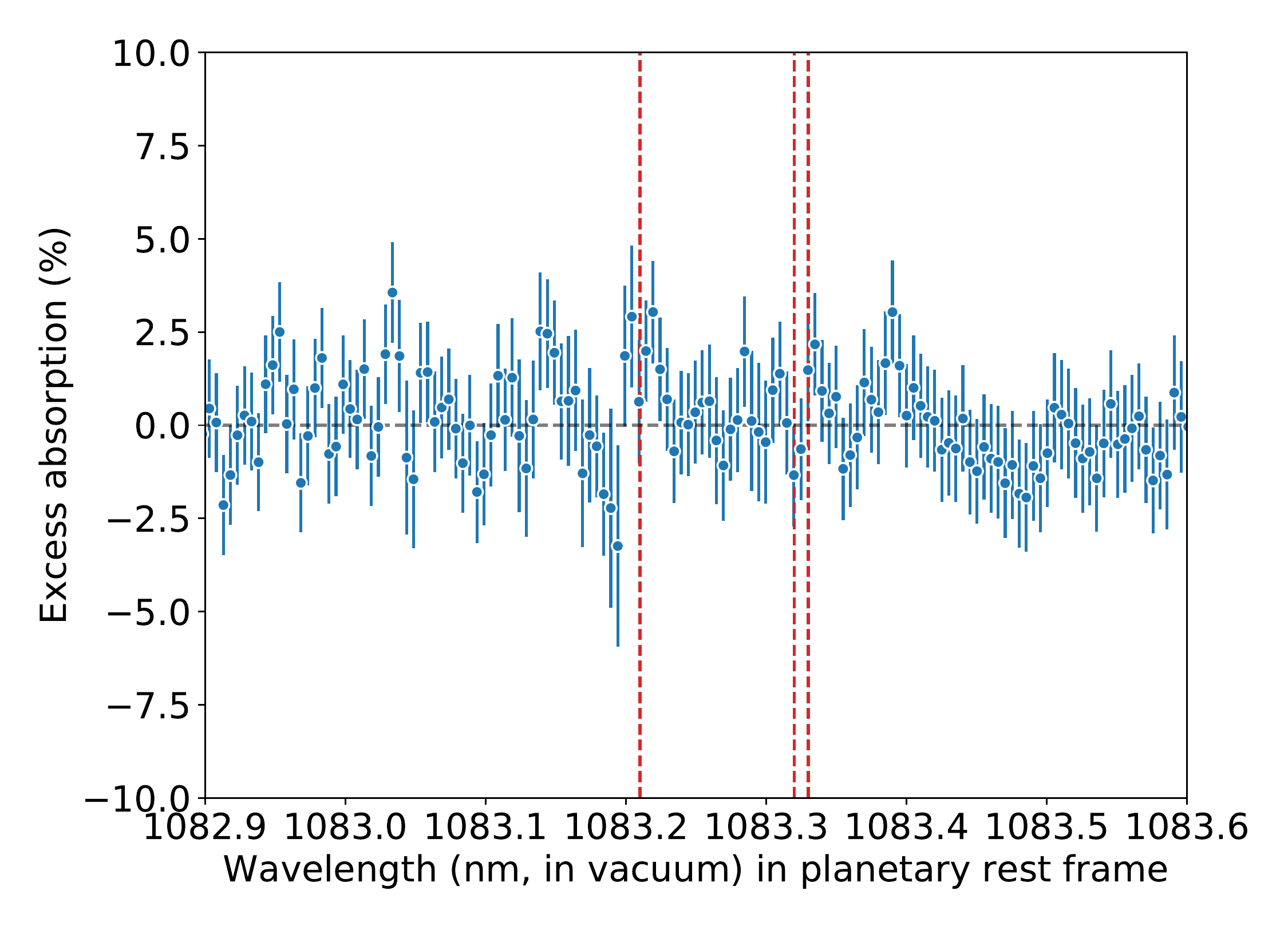}
\caption{Transmission spectrum of WASP-127~b around the He triplet. Absorption is positive.}
\label{transm_spec}
\end{figure}

The time series of the ratio between the individual spectra obtained during the observation and the master out-of-transit spectrum is shown in Fig. \ref{time_series}. The horizontal structure between phases $-0.02$ and $-0.01$ in Fig. \ref{time_series} is likely due to normalization issues caused by the lower S/N of these exposures. Based on the technique of \citet{2015A&A...577A..62W}, we computed an individual transmission spectrum as $f_{t, k} = 1 - f_\mathrm{in, k} / F_\mathrm{out}$, where $f_\mathrm{in, k}$ is the kth in-transit spectrum and $F_\mathrm{out}$ is the master out-of-transit spectrum. We note that $f_\mathrm{in, k}$ was calculated by shifting the spectrum of WASP-127 to the planetary rest frame. The final transmission spectrum $F_t$ was computed by combining the individual ones as a median, and the result is shown in Fig. \ref{transm_spec}.

The average precision of the transmission spectrum of WASP-127~b is 1.9\% per wavelength bin. This level of uncertainty is comparable to the one obtained for one transit of WASP-107~b with the CARMENES spectrograph \citep{2019A&A...623A..58A}. For comparison purposes, the number of measured spectra in both programs is similar, both WASP-127 and WASP-107 possess similar J magnitudes, but the Gemini South Telescope has a mirror size of 8.1~m, while CARMENES is installed on a 3.5~m telescope.

An inspection of the spectral ratio time series and the transmission spectrum does not reveal a strong planetary absorption in the He triplet. We computed the integrated flux inside the wavelength range [1083.280, 1083.355] nm, the same bandwidth as in \citep{2019A&A...623A..58A}, and found an in-transit excess absorption consistent with null, $0.42\% \pm 0.36\%$, which corresponds to an upper limit of 0.87\% at 90\% confidence. In this project, we used a slit width of 0.25 arcsec (three pixels), but future projects with this instrument could benefit from a higher S/N if the widest slit width were used (0.34 arcsec, or four pixels). In this case, there would be some loss in spectral resolution ($R \sim\ $50,000), but given that the instrument is not stabilized in the first place and that the final error in wavelength calibration is $\sim$1.6 pixels, such a loss is not expected to severely affect how well we are able to trace atmospheric signals in Doppler space for short-period, massive exoplanets.

WASP-127~b and its host star are not as well studied as other systems for which we currently have detected He absorption signals. Using the most up-to-date planetary parameters (Seidel et al., in prep.), and assuming a mean molecular weight of $2.3 m_H$, we calculated the lower-atmosphere scale height $H_{\rm eq}$ of WASP-127~b as $2109^{+313}_{-231}$ km. This is the largest scale height for a planet with a published \ion{He}{I} transmission spectroscopy study to date \citep[see, e.g.,][and references therein]{2019A&A...629A.110A}. The resulting 90\% confidence upper limit for the ratio between the equivalent height of the absorbing atmosphere $\delta_{R_{\rm pl}}$ and the lower atmosphere scale height $H_{\rm eq}$ is 18.77. Converting this upper limit to an estimate of atmospheric escape rate is complicated by the number of unknowns, such as the temperature of the escaping material, the actual extreme ultraviolet (EUV) irradiation that the planet receives, the departure of the extended atmosphere from axial symmetry and from local thermal equilibrium.

\begin{figure}[h]
\includegraphics[width=\hsize]{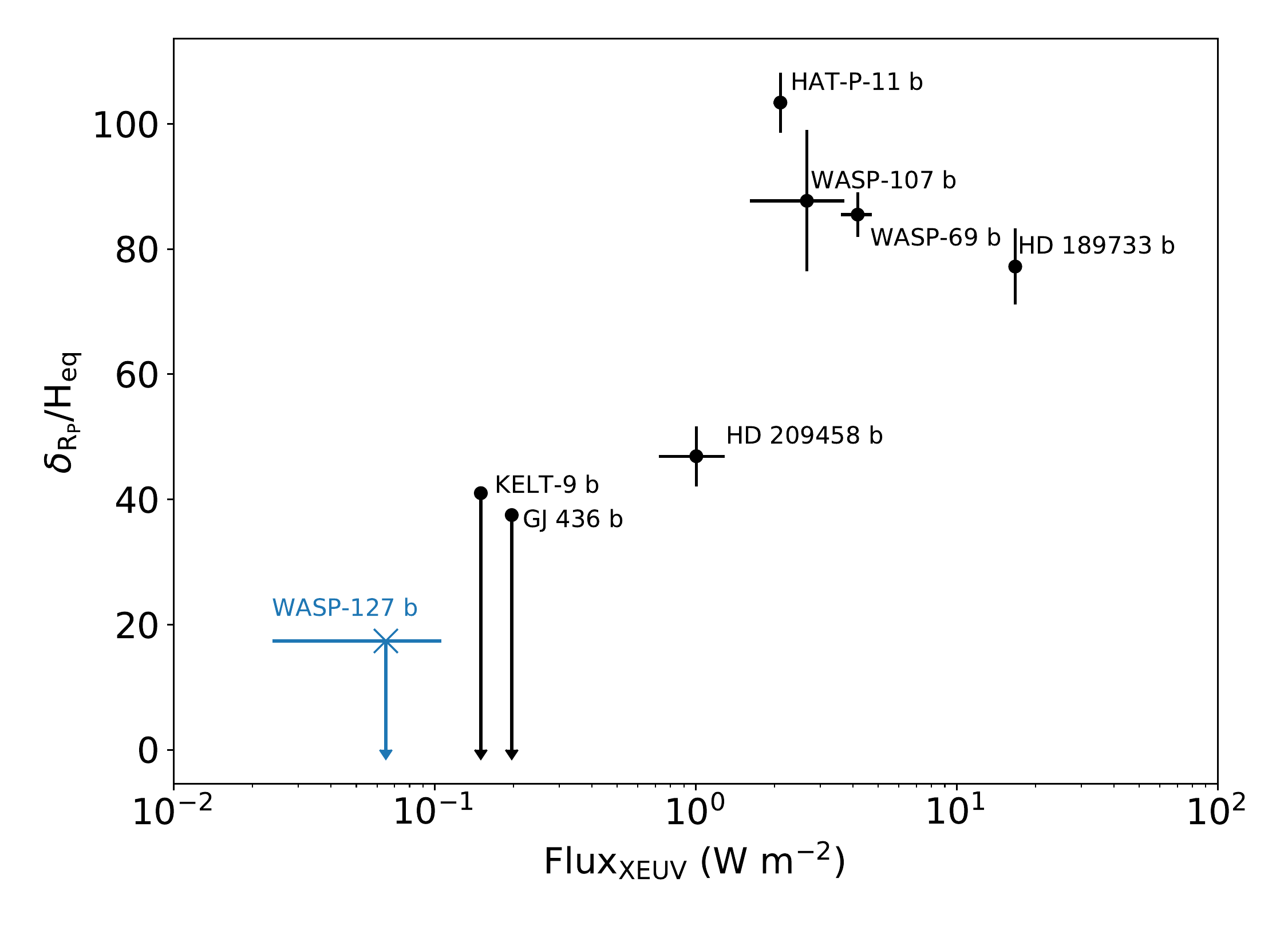}
\caption{\ion{He}{I} ionizing irradiation ($\lambda <$ 504 \AA) versus the equivalent height of the absorbing \ion{He}{I} particles in the upper atmosphere of WASP-127~b. The high-energy irradiation was estimated with two different methods (see main text), which define the limits of the horizontal uncertainty. The data points for other targets were measured by \citet{2018Sci...362.1388N} and \citet{2019A&A...629A.110A}.}
\label{XEUV_vs_delta}
\end{figure}

For comparison purposes, we plotted this result in the high-energy irradiation versus $\delta_{R_{\rm pl}} / H_{\rm eq}$ in Fig. \ref{XEUV_vs_delta}. There are currently no publicly available observations of WASP-127 in X-rays or UV wavelengths, so we needed to estimate the high-energy irradiation of the planet. We used two approaches for this estimate: i) using an SED model of a similar star (HD~330075, same G5-type as WASP-127) from the {\tt X-exoplanets} database\footnote{\footnotesize{Available at \url{http://sdc.cab.inta-csic.es/xexoplanets/jsp/homepage.jsp}.}} \citep{2011A&A...532A...6S}; and ii) using the age and $B-V$ relations from \citet{2012MNRAS.422.2024J} and extrapolating the X-rays luminosity to $\lambda = 504$ \AA. The ionizing flux that populates the metastable \ion{He}{I} triplet is limited to wavelengths that are shorter than 504~\AA\ \citep[XEUV;][]{2019ApJ...881..133O}. These techniques yield XEUV irradiation levels of 0.092 and 0.024 W~m$^{-2}$ at the semi-major axis of WASP-127~b, respectively. Despite possessing a large scale height, one of the explanations for the lack of detectable \ion{He}{I} in the upper atmosphere of WASP-127~b is that the planet receives very little high-energy irradiation compared to planets with a detectable signal. This also seems to be the case for GJ~436~b and KELT-9~b. 

\citet{2017A&A...599A...3L} estimated the age of the WASP-127 system using the isochrones method, which is reliable for old Sun-like stars \citep[see, e.g.,][]{2018MNRAS.474.2580S}, and they derived a value of $11.41 \pm 1.80$ Gyr. The old age of the host star and consequently its relatively weak high-energy output could be part of the explanation as to why there is no detectable He in WASP-127~b. However, the nondetection in the young planet K2-100~b \citep[see the discussion in][]{2020MNRAS.495..650G}, which is under a stronger X-EUV irradiation than WASP-127~b, shows that the atmospheric escape history also affects the detectability of the \ion{He}{I} feature.

\section{Conclusions}

We have presented a pilot study to assess the feasibility of using the Phoenix spectrograph, a visiting instrument on the Gemini South telescope, to detect \ion{He}{I} in transmission around transiting hot-Jupiters. We used a standard procedure to reduce the data to allow for reproducibility. The observed spectra had an average S/N of $\sim$26 near the \ion{He}{I} triplet. In total, we measured 12 exposures of 900~s during the transit of WASP-127~b, ten of which were out of transit and two exposures were discarded because of cloud coverage.

The final transmission spectrum of WASP-127~b near the \ion{He}{I} triplet has an average precision of 1.9\% per wavelength bin. For reference purposes, a transit observation of WASP-107~b using the CARMENES spectrograph rendered a comparable precision for its transmission spectrum; both host stars have a similar J magnitude and were observed for a similar amount of time. This result illustrates the potential to use the Phoenix spectrograph as a viable alternative for this purpose. We do not detect an in-transit excess absorption due to the presence of He, and we constrain it to $0.42\% \pm 0.36\%$ between the wavelength range [1083.280, 1083.355] nm. To date, WASP-127~b has displayed the least amount of atmospheric \ion{He}{I} absorption relative to its scale height when compared to other detections and nondetections. The most likely explanation for this lack of metastable He in the upper atmosphere of WASP-127~b is the low amount of ionizing irradiation (1-504 \AA) from its host star, which in turn can be related to the old age of the system ($11.40 \pm 1.80$ Gyr).

There are two main limitations when using the Phoenix spectrograph to conduct He searches in transmission: i) the lack of an automated pipeline renders the data reduction processes difficult to perform and reproduce; and ii) since this is a visiting instrument, its availability on the Gemini South telescope is dictated by the demand of the community during the first phase of the call for observing proposals. Given the promising results we have presented, we hope that this pilot study encourages the community to apply for observing time using Phoenix and to develop an automated pipeline to reduce its data. Until the date of the observation, this was the only spectrograph available in the Southern Hemisphere with a promising prospect to conduct this type of observation. In the near future, other southern instruments such as CRIRES+ \citep{10.1117/12.2314150} and NIRPS \citep{2017SPIE10400E..18W} will be able to perform such studies; however, having an extra instrument would help distribute the pressure that would otherwise concentrate onto the newer spectrographs.

\begin{acknowledgements}
   LAdS thanks the Gemini Observatory staff and K. Hinkle for the technical support during data analysis, and K. Lam for the discussion on stellar parameters. The authors thank the anonymous referee for the swift and helpful review. This project has received funding from the European Research Council (ERC) under the European Union’s Horizon 2020 research and innovation programme (project {\sc Four Aces}; grant agreement No 724427), and it has been carried out in the frame of the National Centre for Competence in Research PlanetS supported by the Swiss National Science Foundation (SNSF). JM thanks FAPESP, grant number 2018/04055-8. SGS is supported by FCT -- Funda\c{c}\~ao para a Ci\^encia e a Tecnologia through national funds and by FEDER -- Fundo Europeu de Desenvolvimento Regional through COMPETE2020 - Programa Operacional Competitividade e Internacionaliza\c{c}\~ao by these grants: PTDC/FIS-AST/32113/2017 and POCI-01-0145-FEDER-32113; PTDC/FIS-AST/28953/2017 and POCI-01-0145-FEDER-028953. This study is based on observations obtained at the international Gemini Observatory, a program of NOIRLab, which is managed by the Association of Universities for Research in Astronomy (AURA) under a cooperative agreement with the National Science Foundation on behalf of the Gemini Observatory partnership: the National Science Foundation (United States), National Research Council (Canada), Agencia Nacional de Investigaci\'{o}n y Desarrollo (Chile), Ministerio de Ciencia, Tecnolog\'{i}a e Innovaci\'{o}n (Argentina), Minist\'{e}rio da Ci\^{e}ncia, Tecnologia, Inova\c{c}\~{o}es e Comunica\c{c}\~{o}es (Brazil), and Korea Astronomy and Space Science Institute (Republic of Korea). We used the open source software SciPy \citep{2020SciPy-NMeth}, NumPy \citep{5725236}, IPython \citep{4160251}, Jupyter \citep{Kluyver:2016aa}, Astropy \citep{2013A&A...558A..33A}, and Matplotlib \citep{Hunter:2007}.
\end{acknowledgements}

\bibliographystyle{aa}
\bibliography{biblio.bib}

\end{document}